\documentstyle[11pt,epsf]{article}

\def\half{{\scriptstyle {1\over 2}}}
\def\ltap{\mathrel{
    \raise.55ex\hbox{$<$}\mkern-14.3mu\raise-.6ex\hbox{$\sim$}}}
\def\gtap{\mathrel{
    \raise.55ex\hbox{$>$}\mkern-14.3mu\raise-.6ex\hbox{$\sim$}}}
\def\into{\mathbin{\to}}

\def\stop{{\tilde t_1}}
\def\stoph{{\tilde t_2}}
\def\zino{{\widetilde Z_1}}
\def\wino{{\widetilde W_1}}
\def\ttbar{{t\bar{t}}}
\def\stoppair{{\stop\,\bar{\!\tilde t}_1}}
\def\met{/\!\!\!\!E_T}

\def\bft{{\cal B}_{t\into\stop}}
\def\bfw{{\cal B}_{\wino\into\ell}}
\def\bfW{{\cal B}_{W\into\ell}}

\def\xtt{\sigma_{\ttbar}}
\def\xts{\sigma_{t\into\stop}}
\def\xss{\sigma_{\stoppair}}

\def\xttss{\sigma_{\ttbar\into\stoppair}}

\def\ets{\varepsilon_1}
\def\ess{\tilde\varepsilon}
\def\ett{\varepsilon_2}
\def\ettss{\ett}
\def\evis{\eta}

\def\LL{{\ell^+\ell^-}}
\def\lb{{\ell\!+\!b}}
\def\pspace{m_\stop \times m_\wino}
\def\pb{{\rm \ pb}}

\def\lnmt{{\cal N}_t}
\def\lnmb{{\cal N}_b}
\def\lnme{{\cal N}_\epsilon}
\def\lognormt{{\cal N}(\bar t,\delta t/t)}
\def\lnorms{{\cal N}(4.05,49.5\%)}
\def\lnormt{{\cal N}(271.5,50.2\%)}
\def\lnorme{{\cal N}(.034,.009/.034)}

\title{
    \LARGE\bf Light 
	$\stop\to\wino b$  
    models are not ruled out by the Tevatron top experiments
    }

\author{
    {\Large John Sender}
    \thanks{
	{\tt sender@uhheph.phys.hawaii.edu}
	}
    \\
    \\{\normalsize\em Department of Physics and Astronomy,}
    \\{\normalsize\em University of Hawaii, Honolulu, HI 96822 USA}
    }
\date{}

\begin{document}
\maketitle

\begin{picture}(0,0)
\put(315,240){UH-511-843-96}
\put(315,225){hep-ph/9602354}
\put(315,210){February, 1996}
\end{picture}
\vspace{-36pt}

\begin{abstract}
SUSY events in which a light stop decays to a chargino may manage
to pass the cuts designed for top counting experiments at the
Tevatron with no significant diminution of signal.  Events in which
a top decays via stop are detected with fair efficiency compared to
SM top events when
$\stop\into b\wino$.  
More importantly, direct pair production of light stops contributes
significantly to the cut samples.
Many SUSY models with stops lighter than 100 GeV
are viable even for 
$\Gamma(t\into\stop) \approx \Gamma( t\into bW)$.  
\end{abstract}

\bigskip
\noindent


Attention has recently focussed on SUSY models with light stops 
  ($\stop$)  
and charginos 
  ($\wino$).  
Many authors \cite{Many} have pointed out that in such light
SUSY scenarios one must check that tops are not lost through
exotic decays to an extent that conflicts with the
Tevatron top counting experiments.
This is certainly a consideration for stops that decay
directly to the LSP 
  ($\zino$)  
via 
  $\stop\into c\zino$,  
where $b$-jets  
and isolated hard leptons are missing in the final state.
The study presented here indicates, however, that light stops
decaying through a chargino are generally not yet constrained
by the Tevatron data.


Light stops decay in one of two ways \cite{HK,BDGGT}.
In the case 
  $\stop\into\zino c$,  
most SUSY 
  $\ttbar$  
events will be lost by the counting experiments since 
  $t\into\zino\stop\into\zino\zino c$  
gives neither a $b$  quark nor a hard isolated lepton.
We do not consider this mode in the present study.
Rather, we look at the other case, 
  $\stop\into\wino b$,  
which dominates whenever kinematically allowed.
The SUSY decay 
  $t\into\zino\stop\into\zino\wino b\into\zino\zino\bar ff'b$  
adds a neutralino pair to the final state of 
  $t\into W b\into\bar ff'b$  
and thus softens the $\ell$ and $b$ tags,
so the counts are suppressed in this case as well.
But both tags are still available
and clearly  less signal is lost than in the 
  $\stop\into\zino c$  
case.  Moreover, the direct stop production final state
  $\stoppair\into b\ell\bar b\ell'\nu\nu\zino\zino$  
also just adds a neutralino pair to 
  $\ttbar\into b\ell\bar b\ell'\nu\nu$,  
so all four potential tags are present and such events can
contribute to the top cut samples.
Similar considerations apply to the $1-$lepton+$b$-jet channel.
For light stops, the efficiency to pass the top selection cuts
is small, but production cross-section is high
(75 GeV stops have 10 times the production cross-section of
175 GeV tops).
The net effect is that what the counting experiments
lose to exotic decays in 
  $\ttbar$  
events, they may make up (and more) by admitting 
  $\stoppair$  
events.


If 
  $t\into Wb$  and 
  $t\into\zino\stop$  
are the only top decays with appreciable branching fraction and
sleptons are all heavy and degenerate in mass and 
  $\stoph$  
is sufficiently heavy, then the top-stop system at the Run I
Tevatron is specified by 6 parameters: the four masses 
  $m_t$,  
  $m_\stop$, 
  $m_\wino$ and 
  $m_\zino$,  
and two branching fractions 
  $\bft$  and 
  $\bfw$.  
In this study we fix 
  $m_t = 180\pm15$  
GeV \cite{Mtop}.
For the case studies shown in the figures, we impose
the familiar relation 
  $m_\zino = \half m_\wino$,  
which is a standard feature of SUGRA models.
Later, to see how sensitive our results are to this assumption,
we quote results for a
  ``light $\zino$"  
case where we fix
  $m_{\zino}=20$ GeV,
and a
  ``heavy $\zino$"  
case where
  $m_{\zino}=m_{\wino}-10$ GeV.
The
  ``heavy $\zino$"  
case arises in models which set the SUSY Higgs mass 
  $|\mu|$  
much smaller than the soft-SUSY-breaking
  {\it SU}$(2)$  
gaugino mass in order to produce a higgsino-like 
  $\wino$  
giving Yukawa strength to the 
  $b\wino\stop$  
coupling.


Finally, we set the leptonic branching fraction of the chargino to
that of the $W$,  
  $\bfw = \bfW \approx 1/9$. {}
In general SUSY models 
  $\bfw$  
can be anywhere in the permissible range 0 to 1/3.
Even the restrictive SUGRA class includes models spanning this whole
range for most 
  $m_\wino \ltap 100$ GeV.
But when squark and slepton masses get heavy, 
  $\bfw$  
generally approaches $1/9$,  so we take it as a representative value.
For 
  $\bfw$  
not too different from this our results scale to other values --
the single lepton 
  $\lb$  
channel goes like 
  $\bfw$,  
and the dilepton like 
  $(\bfw)^2$;  
substantial differences would occur if 
  $\bfw$  
deviated far from $1/9$.  
We also use simple scaling for the SUSY top branching fraction
  $\bft$  
so the parameter space for this study is just 
  $\pspace$.  


We use ISAJET version 7.14 \cite{Isajet}
with the CTEQ2L \cite{CTEQ2L} structure functions
to estimate the SUSY contribution to the top counting experiments.
We model the experimental conditions at the Tevatron by
incorporating a toy calorimeter with segmentation
  $\Delta\eta\times\Delta\phi = 0.1\times 0.087$  
and extending to $|\eta | = 4$  
in our simulation. We have assumed an energy resolution of 
  $50\% /\sqrt{E}$ ($15\% /\sqrt{E}$)  
for the hadronic (electromagnetic) calorimeter. Jets are
defined to be hadron clusters with minimum $E_T$  
as given in the text, within a cone of 
  $\Delta R  = \sqrt{\Delta\eta^2 +\Delta\phi^2} = 0.7$  
and $|\eta_j| < 3.5$.  
We consider an electron (muon) to be isolated if 
  $p_T(e) > 8$  GeV 
  ($p_T(\mu) > 5$  GeV)
and the hadronic energy in a cone with 
  $\Delta R = 0.4$  
about the lepton does not exceed
  $\min$($\frac{1}{4}E_T(\ell), 4$ GeV).
Non-isolated electrons are included as part of the accompanying
hadron cluster.


In our calculations, we use CDF-inspired cuts \cite{CDFbig}
to compare with the data in the 1-lepton+$b$-tag  
  ($\lb$)
and dilepton 
  ($\LL$)  
channels.  The 1-lepton+$b$-tag  cuts are:
(a) 
      $\met > 20$ GeV,
(b)
    an isolated lepton with 
      $p_T > 20$  GeV and 
      $|\eta|<1$,  
(c)
    no dilepton with 
      $70 < m_{\LL} < 110$ GeV,
(d) 
    $B$  hadron with $p_T > 15$ GeV and 
      $|\eta|<2$  
    (tagging efficiency of 20\% per taggable $B$),  
and
(e)
    at least 3 jets with $p_T > 15$  GeV and $|\eta|<2$.  
In the dilepton channel, we require
(a) 
      $\met > 25$  GeV,
(b)
    an isolated lepton with $p_T > 20$  GeV and $|\eta|<1$,  
(c)
    an opposite sign lepton with $p_T > 20$  GeV,
(d)
    no dilepton with $75 < m_{\LL} < 125$  GeV,
(e)
    at least 2 jets with $p_T > 10$  GeV and $|\eta|<2$,  
 and
(f)
    if 
      $\met < 50$ GeV,
    then no jet or lepton closer to 
      $\met$  
    than 20 degrees in azimuth.


For each of the two channels, we generate four sets of events
which are subjected to our simulation cuts:
(a) Standard Model 
  $\ttbar$  
 events,
(b) 
  $\ttbar$  
 events in which one of the tops decays to 
  $\stop$,  
(c) 
  $\ttbar$  
 events in which both tops decay to 
  $\stop$,  
 and
(d) direct 
  $\stoppair$  
 production events.
Calling the cross-sections after cuts for these four sets as 
  $\xtt$,  
  $\xts$, 
  $\xttss$ and 
  $\xss$,  
respectively, then we calculate for each point in our parameter
space the quantities 
  $\ets=\xts/\xtt$,  
  $\ettss^2=\xttss/\xtt$ and 
  $\ess=\xss/\xtt$. 
By normalizing in this way, we reduce the sensitivity of our
results to simulation differences.
  $\ets$  
 ($\ettss$)
is the relative efficiency for counting an event with one (two)
SUSY top decays.
  $\ess$  
represents the relative enhancement of the count due to direct
  $\stoppair$  
production.


\begin{figure}
\begin{center}
\leavevmode
\makebox[0pt]{
\epsffile{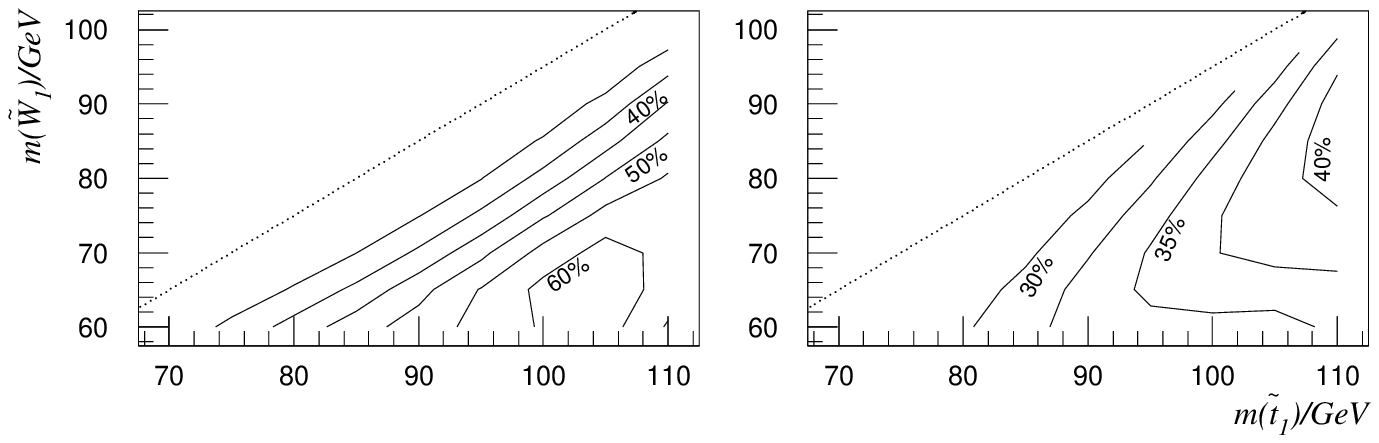}}
\raisebox{1.2in}{
\makebox[0pt][l]{
\hspace{-2.0in}
  \makebox[0pt]{$\ets^{\lb}$}  
\hspace{2.67in}
  \makebox[0pt]{$\ets^{\LL}$}  
}}
\end{center}
\caption{
    Relative efficiency for counting 
      $\ttbar$  
    events with a SUSY decay 
      $t\into\zino\stop\into\zino\wino b$.  
    The left contour plot shows results for the 1-lepton+$b$-jet  
    channel, and the right for the dilepton channel.
    Calculations assume 
      $m_t = 180$ GeV,
      $m_{\zino} = m_{\wino}/2$ and 
      $\bfw=\bfW$.}  
\label{FIGone}
\end{figure}


Figure 1 shows contours of 
  $\ets$,  
the efficiency for counting events with a single SUSY top decay,
plotted in the 
  $\pspace$  
plane.  Substantial fractions of such events pass the top cuts
in either channel (although this does not mean that the SUSY stop
signal is detectable {\em above} the top background \cite{BDGGT}).
Recall that we have set 
  $m_{\zino}=\half m_{\wino}$  
for Fig. 1.  We also performed calculations in this plane for a
  ``light $\zino$"  
case using 
  $m_{\zino}=20$ GeV,
where we find that the efficiencies increase in both channels by
about 30\% of their values (i.e., 30\% goes to 40\%).
Further, we have looked at a
  ``heavy $\zino$"  
case where we fix 
  $m_{\zino}=m_{\wino}-10$  GeV.
Here, the lepton+$b$-jet  efficiencies fall by a half and the
dilepton efficiencies all drop to less than 10\%.
We do not show figures for 
  $\ettss$:  
for lepton+$b$-jet  we get 
  $\ettss/\ets \approx 70\%$  to 90\%,
and in the dilepton channel 
  $\ettss/\ets \gtap 90\%$.  


\begin{figure}
\begin{center}
\leavevmode
\makebox[0pt]{
\epsffile{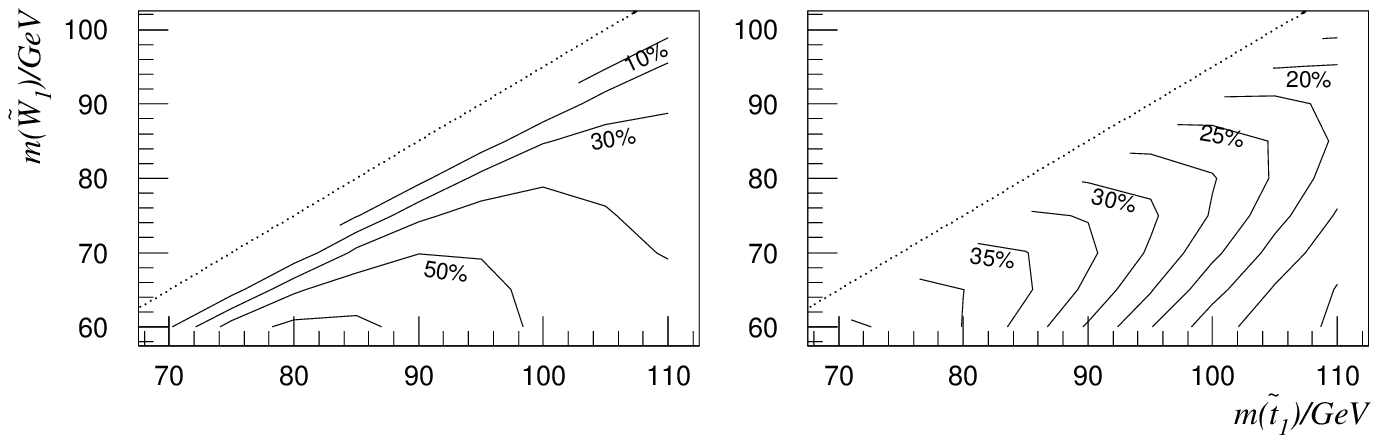}}
\raisebox{1.2in}{
\makebox[0pt][l]{
\hspace{-2.0in}
  \makebox[0pt]{$\ess^{\lb}$}  
\hspace{2.67in}
  \makebox[0pt]{$\ess^{\LL}$}  
}}
\end{center}
\caption{
    Enhancement of the 
      ``$\ttbar\,$"  
    signal due to direct 
      $\stoppair$  
    production.  SUSY model assumptions are as in Figure 1.}
\label{FIGtwo}
\end{figure}


Figure 2 shows the results for 
  $\ess$,  
the enhancement due to direct stop production.  Again, there is
appreciable SUSY contribution.  In the lepton+$b$-jet  channel, 
  $\stoppair$  
with stops lighter than 100 GeV adds about 40\% for stops away
from the kinematical limit
  $m_\stop \ltap m_\wino + m_b$.  
This enhancement practically vanishes at the limit since both
$b$'s come from 
  $\stop\into\wino b$  
and hence neither is hard enough to tag.
For dileptons, the enhancement is generally 
  $\gtap$ 25\%,
softening as the chargino becomes too light to produce hard enough
leptons.  In both channels, the effect is greater for lighter stops
since the increasing production cross-section overcomes the
decreasing efficiency.
In the
  ``light $\zino$"  
case, the enhancement almost doubles in the lepton+$b$-jet  
channel and adds another 50\% in the dilepton channel.
The enhancement in both channels disappears for heavy 
  $\zino$'s,  
dropping to only a few percent when 
  $m_{\zino}=m_{\wino}-10$ GeV.



\begin{table}[b]
\begin{center}
\begin{tabular}{|l|c|c|c|} \hline
    &
    & $\lb$ channel
    & $\LL$ channel
    \\ \hline \hline
integrated luminosity
    & ${\cal L}$  
    & $67\pm 5 \pb^{-1}$
    & $67\pm 5 \pb^{-1}$
    \\ \hline
cut efficiency
    & $\epsilon$  
    & $0.034\pm 0.009$
    & $0.0078 \pm 0.0008$
    \\ \hline
estimated background
    & $b$  
    & $5.5\pm 1.8$
    & $1.3\pm 0.3$
    \\ \hline
\# of observed events
    & $N$  
    & 21
    & 6
    \\ \hline
\end{tabular}
\end{center}
\caption{CDF data.}
\end{table}


To constrain branching fractions, we have to predict counts.
Table 1 gives the CDF values we will use \cite{CDFsummer}.
They are based on 
  $67 \pb^{-1}$  
of Run I data.  For a given experiment, we expect
  $s=\sigma\cdot {\cal L}\cdot \epsilon\cdot\evis$  
counts, where
  $\sigma=$  $\ttbar$
production cross-section,
  ${\cal L} =$  
integrated luminosity,
  $\epsilon =$  
efficiency after detection and cuts, and 
  $\evis$  
encodes the possible effects of non-SM physics
  ($\evis = 1$  
in the Standard Model).
We parameterize the cross-section as
  $\sigma = 4.75 e^{(175-m_t)/31.5}(1\pm 15\%)$  
 \cite{CMNT}
(15\% is the QCD theory uncertainty)
and suppose that it has a log-normal distribution with mean 
  $\sigma$  
and variance
  $(\delta\sigma/\sigma)^2 = (\delta m_t/31.5)^2 + (15\%)^2$,  
which for 
  $m_t=180\pm15$ GeV are
  $\sigma = 4.05$ pb and
  $\delta\sigma/\sigma = 49.5\%$.  
We write this distribution as 
  $\lnorms$.
Together with the luminosity from Table 1, we expect the number of
produced 
  $\ttbar$  
events, 
  $t=\sigma{\cal L}$,  
to be distributed as
  $\lnmt\equiv\lognormt = \lnormt$.
The efficiencies come from Table 1; in the 
  $\lb$  
channel we use
  $\lnme = \lnorme$,
and similarly for the 
  $\LL$  
channel.  The backgrounds $b$  from Table 1 are also distributed
in this fashion.  We compute the lower limit for 
  $\evis$  
at the 95\% confidence level by solving
  $ 5\% =
    \int\! d\ln t\, \lnmt
    \int\! d\ln \epsilon\, \lnme
    \int\! d\ln b\, \lnmb
    \,{\cal P}_N $,
where
  $ {\cal P}_N =
	\left(
	e^{-(b+s)} \sum_{n\geq N}{(b+s)^n / n!}
	\right)\left(
	e^{-b} \sum_{n\leq N} {b^n / n!}
	\right)^{-1} $
is the probability of getting $N$  or more signal+background
events given no more than $N$  background events,
with Poisson parameters 
  $s=t\epsilon\evis$  
for signal and $b$  for background
A similar calculation yields the limit on 
  $\evis$  
for the
  $\LL$  
channel.  Finally, we get a combined limit on 
  $\evis$  
by solving
  $ 5\% = \int\! d\ln t\, \lnmt
    \left[ {\cal C}_{\lb}{\cal C}_{\LL}
	( 1-\ln {\cal C}_{\lb}{\cal C}_{\LL})\right] $
where
  ${\cal C}_{\lb}=\int\! d\ln \epsilon\, \lnme
    \int\! d\ln b\, \lnmb \,{\cal P}_N$
using the 
  $\lb$  
values for $\epsilon$,  $b$ and $N$,  and likewise for 
  $\LL$  
\cite{Eadie}.


In terms of the branching fraction 
  $\bft$  
and the relative efficiencies, the parameter 
  $\evis$  
is
  $ \evis =
    (1-\bft)^2
    + 2(1-\bft)(\ets\bft)
    + (\ett\bft)^2
    + \ess $.  
Setting 
  $\ets=\ett=\ess=0$  
in both channels, we get
  $\bft<22\%$  ($\lb$), 
  $\bft<27\%$  ($\LL$) and
  $\bft<18\%$  (combined).
This is the usual formulation for setting
branching limits on ``invisible" new decay modes, such as
  $t\into\zino\stop$  
where 
  $\stop\into\zino c$.  
It actually underestimates the limit, since one expects a
finite value for 
  $\ets$  
in the 
  $\lb$  
channel.  That is, if 
  $\bar t\into\zino\zino\bar c$  
while 
  $t\into \ell\bar\nu b$,  
then $t$  can produce an identifiable $b$  and lepton while 
  $\bar t$  
gives a $c$-jet  plus a contribution to 
  $\met$.  
Such an event can pass the cuts.  Roughly, one expects
  $\ets \ltap 1/4$  
since there are only half as many available $b$'s  
and half as many available hard isolated leptons in the final state.
We have examined a variety of 
  $\stop\into\zino c$  
cases with our simulation and find 
  $\ets$  
in the range $16-23\%$.  
Taking 
  $\ets=20\%$,  
the branching fraction limit as calculated above weakens slightly to 
  $\bft<20\%$  
(combined) \cite{Mrenna}.


\begin{figure}
\begin{center}
\leavevmode
\makebox[0pt]{
\epsffile{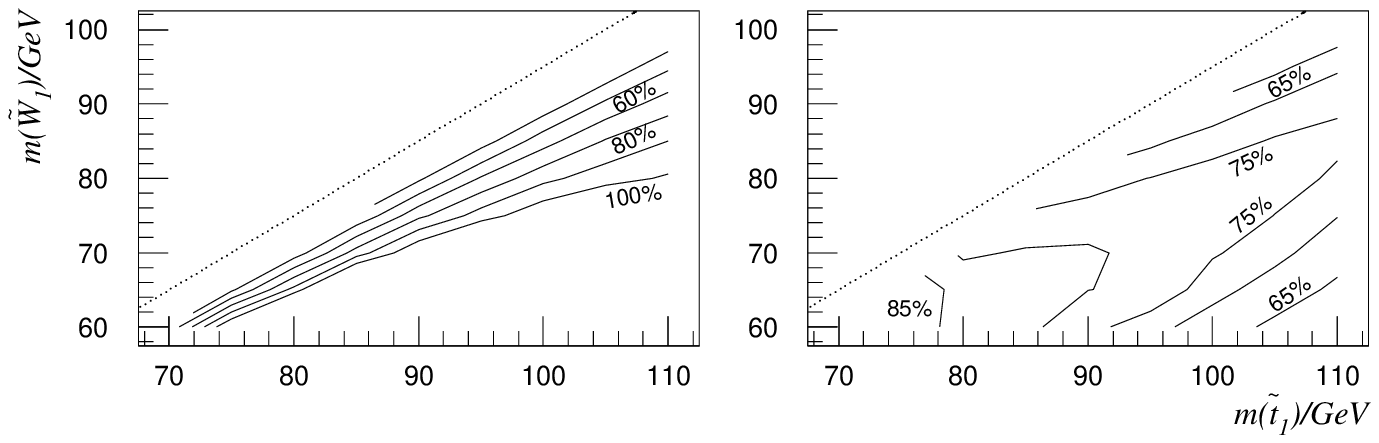}}
\raisebox{1.2in}{
\makebox[0pt][l]{
\hspace{-1.85in}
  \makebox[0pt]{$\bft^{\,\lb}$}  
\hspace{2.67in}
  \makebox[0pt]{$\bft^{\,\LL}$}  
}}
\end{center}
\caption{
    95\% confidence level upper limits on the
      $t\into\zino\stop\into\zino\wino b$  
    branching fraction.  Model assumptions are as in Figure 1.}
\label{FIGthree}
\end{figure}


For the 
  $\stop\into\wino b$  
models of interest in this study, we can use our calculated values of 
  $\ets$,  
  $\ett$ and 
  $\ess$  
to get branching fraction limits.  These limits are plotted in
Figure 3, again under the assumptions
  $\bfw=\bfW$  
and 
  $m_\zino=\half m_\wino$.  
The lepton+$b$-jet  channel only offers branching fraction
contraints as the kinematical limit is approached -- none of the
ample parameter space below the 100\% contour is ruled out by
this experiment for any 
  $\bft$.  
The dilepton channel is less accomodating, but for stops 
  $\ltap$  
100 GeV most models with 
  $\bft<3/4$  
are still viable.  We do not show the branching fraction limits
for the two channels combined, but they generally follow the 
  $\lb$  
limits above the 
  $\lb$ 100\%
line, and the 
  $\LL$  
limits below this line.  In the
  ``light $\zino$"  
case, 100\% branching fractions are allowed over the whole plane.
For the
  ``heavy $\zino$"  
case, the lepton+$b$-jet channel has 
  $\bft$  
limits of 
  $\ltap 40\%$,  
while the dilepton channel gets little help from SUSY: 
  $\bft \ltap 27\%$,  
approaching the 20\% limit for ``invisible" decays we calculated
above.


The main determinant of the branching fraction limits we report here
is the lower limit on $m_t$.  When tops are allowed to be light
(relative to the mass one would estimate by rate), then they can be
produced in excess and the excess hidden by exotic decays.
If the mass uncertainty were to decrease while preserving a high
central value, then tighter limits could be deduced.
For instance, if we replace $180\pm15$  by $180\pm10$ in our
analysis, then the 80\% contours in Figure 3 would become 65\%
contours.  
Thus it seems unlikely that inclusion of {\em all} the Tevatron Run I
data from both CDF and D0 can produce much more confining limits on 
  $\bft$  
than we have exhibited here.
In Run II, when enough integrated luminosity is gathered to
substantially improve 
  $\delta m_t$,  
a direct search for
  $p\bar p\into \stoppair X\into b\bar b \wino\bar\wino X$  
becomes feasible over most of the parameter space we have investigated
in this study \cite{BST}.
The direct search will generally be more probative than
branching-fraction analyses for 
  $\stop\into\wino b$  
models with 
  $\bft \ltap 50\%$.  
This is so because if there is a sizeable SUSY component to the
top cut samples, then one would expect to see this in the soft
parts of the distributions of kinematical quantities.
This is particularly the case for the
  $\stoppair$  
enhancement, which comes from the tails of the
  $\stoppair$  
distributions.


Given the large enhancements exhibited in Figure 2, one might suppose
that some models are ruled out because they actually produce
{\em too much} signal, rather than too little.
Such effects would vanish, though, if $m_t$  is taken to its upper
experimental limit, due to the resulting decrease in 
  $\xtt$  
and increase in
  $\bft$.  
Indeed, from an analysis as above we can get the {\em upper}
95\% confidence level limits
  $\evis < 4.3$  
for 
  $\lb$  
and 
  $\evis < 5.9$  
for 
  $\LL$.  
The stop models we have investigated here are unlikely to produce
such excesses for any values of 
  $\bft$  
or 
  $\bfw$.  


In summary, it appears that many light 
  $\stop\into\wino b$  
scenarios cannot be excluded by the Tevatron counting experiments,
even if
  $\Gamma(t\into\stop)$  
is comparable to the Standard Model width
  $\Gamma( t\into bW)$.  
An important exception is models with a small 
  $\wino-\zino$  
mass difference, which require 
  $\bft\ltap25\%$  
since they fail to generate hard leptons.
Branching fraction limits also become tighter when 
  $\bfw \ll \bfW$,  
which is also depleted in leptons, or when 
  $m_{\wino}+m_b$  
approaches 
  $m_{\stop}$,  
softening the $b$'s.  
Most SUGRA 
  $\stop\into\wino b$  
models remain viable.
By the time sufficient statistics are available in the top sector to
significantly improve on the branching fraction limits reported here,
direct 
  $\stoppair$  
searches should have confirmed or ruled out these scenarios.


The author appreciates many useful conversations with X. Tata,
and wishes to thank H. Baer for reading he manuscript.
This work was supported in part by
a grant from the U. S. Department of Energy,
DE-FG-03-94ER40833.


\end{document}